\documentclass[12pt,thmsa]{article}
\usepackage{sw20aip}

\input{tcilatex}
\begin{document}

\author{Preston Jones, California State University Fresno \and Lucas Wanex,
University of Nevada Reno}
\title{Coordinate Conditions for a Uniformly Accelerated or Static Plane Symmetric
Metric}
\date{December 13, 2004}
\maketitle

\begin{abstract}
The coordinate conditions for three exact solutions for the metric
components of a coordinate system with constant acceleration or of a static
plane symmetric gravitational field are presented. First, the coordinate
condition that the acceleration of light is constant is applied to the field
equations to derive the metric of a coordinate system of constant
acceleration. Second, the coordinate conditions required to produce the
metrics of Rindler and Lass are applied to the field equations to calculate
the components of these two metrics and the coordinate velocities and
coordinate accelerations for light of these two metrics are compared to the
coordinate system of constant acceleration.
\end{abstract}

\section{Introduction}

In developing the modern theory of gravitation Einstein \cite{Einstein 45}%
\cite{Einstein et al} assumed that constant acceleration is
indistinguishable from a uniform gravitational field. This assumption lead
to the equivalence principle and ultimately to a geometric theory describing
the kinematics of these indistinguishable systems. In this geometric theory
the principle mathematical object for the description of particle
trajectories is the metric, which defines the components of the 4-space
invariant interval. There are at least three compelling reason for the
continuing interest in the study of the relation between constant
acceleration and a uniform field. First, the accelerated system is the most
general form of motion for real physical systems. Second, the accelerated
system is the historic basis of the modern theory of gravity. Finally, a
good understanding of the accelerated system informs a broader understanding
of general relativity as recognized by Misner et al \cite{Misner et al} when
they observed that ''It will be helpful in many applications of gravitation
theory''.

Due to the central importance of the accelerated system for the
understanding of general relativity, studies of the accelerated system are
well represented in the physics literature\cite{Lass}\cite{Rindler 66}\cite
{Rohrlich}\cite{Tilbrook}\cite{Tolman}. However, there is no general
agreement on the exact form of the metric components of an accelerated
coordinate system. The reason for the appearance of different forms of the
metric for an accelerated coordinate system is that the field equations do
not uniquely determine the metric components. The problem of developing
sufficient restrictions on the solution to the field equations in order to
exactly determine the components of the metric is not peculiar to the static
plane symmetric field. It is generally necessary to impose some additional
restrictions on the components of the metric separate from the physical
constraints. These restrictions are referred to as coordinate conditions and
these together with the physical constrains uniquely determine the metric.
Three different sets of coordinate conditions will be considered here for
the field equations of a static and plane symmetric gravitational field. The
first metric will be calculated by imposing the coordinate condition that
the coordinate acceleration of light is a constant. This will be compared to
the previously calculated metrics of Lass \cite{Lass} and Rindler \cite
{Rindler 66}\cite{Rohrlich} and the corresponding coordinate conditions for
these two metrics.

In order to determine the metric components of a static plane symmetric
field or equally an accelerated coordinate system the field equations are
solved, resulting in a relation between the metric components but not a
unique solution. The solution to the field equations is also compared to the
requirement that the metric of an accelerated coordinate system be conformal
flat. The solution of the field equations for a static plane symmetric field 
\cite{Rohrlich} is found to provide a relation between the metric components
that is indistinguishable from the conditions for a conformally flat metric.%
\cite{Tilbrook} This equivalence of the solution to the field equations and
of a conformally flat metric is a formal demonstration of Einstein's
equivalence principle. Finally, by applying appropriate coordinate
conditions, the time and 3-space components of the metric are uniquely
determined and the coordinate time is found to be the same for all
coordinate conditions.

\section{Coordinate Conditions and the Metric}

The metric for any 4-space has 16 components which are in general
independent. Of these 16 components 6 are uniquely determined by the field
equations\cite{Weinberg}. The complete determination of all 16 components
requires the application of both physical restrictions on the metric and on
the coordinates of the metric. The restrictions on the coordinates for the
metric components are referred to as coordinate conditions. There are a
great variety of possible coordinate conditions that can be applied to
uniquely determine the metric components. Three will be considered here for
the solution of the field equations for a static plane symmetric field.
These coordinate conditions are that the coordinate acceleration of light is
a constant, the application of the harmonic coordinate conditions, and the
requirement that the coordinate 3-space interval is the same as the proper
3-space interval. The application of the later two coordinate conditions
result in the Lass and Rindler metrics respectively.

The equivalence principle requires that the metric for an accelerating
coordinate system is the same as that of a uniform and constant
gravitational field. The metric of a uniform field will be static and
unchanged by coordinate transformations in any plane perpendicular to the
acceleration. Under these conditions all derivatives in the field equations
are zero except parallel to the acceleration and in order to be consistent
with a special relativistic momentarily co-moving frame (MCMF) we also
require that $\frac{\partial g_{11}}{\partial z}=$ $\frac{\partial g_{22}}{%
\partial z}=0$. Under these restrictions the metric may be written, in plane
symmetric or Cartesian coordinates, 
\begin{equation}
ds^{2}=-c^{2}d\tau ^{2}=g_{00}dt^{2}+g_{11}dx^{2}+g_{22}dy^{2}+g_{33}dz^{2},
\end{equation}

\noindent or more conveniently in a system of units where $c=1$, 
\begin{equation}
ds^{2}=-V^{2}dt^{2}+dx^{2}+dy^{2}+U^{2}dz^{2}.
\end{equation}

\noindent It is noteworthy that in this system of units the acceleration due
to gravity at the surface of the earth is $g\simeq 1$.

This metric must as well match to first order $g_{00}\simeq -\left( 1+2%
\mathbf{a}\cdot \mathbf{z}\right) $, where $\mathbf{a}$ is the acceleration
from the weak field approximation, $g_{00}\simeq -\left( 1+2\phi \right) $.
Here $\phi $ is the potential of the Newtonian solution for a static plane
symmetric field.

\section{The Field Equations and the Equivalence Principle}

In order to calculate the metric of a plane symmetric field, the field
equations must be solved,

\begin{equation}
R_{\mu \nu }=-8\pi G\left( T_{\mu \nu }-\frac{1}{2}g_{\mu \nu }T_{\lambda
}^{\lambda }\right) ,
\end{equation}

\noindent where $R_{\mu \nu }=0$ in a source free region. The RHS of this
equation is determined by the mass-energy source terms and the LHS by the
4-space geometry. The components of the Ricci tensor are,

\begin{equation}
R_{\mu \nu }=-\frac{\partial }{\partial x^{\alpha }}\Gamma _{\mu \nu
}^{\alpha }+\Gamma _{\mu \beta }^{\alpha }\Gamma _{\nu \alpha }^{\beta }+%
\frac{\partial }{\partial x^{\nu }}\Gamma _{\mu \alpha }^{\alpha }-\Gamma
_{\mu \nu }^{\alpha }\Gamma _{\alpha \beta }^{\beta },
\end{equation}

\noindent and the Christoffel symbols in this equation are,

\begin{equation}
\Gamma _{\mu \nu }^{\sigma }=\frac{1}{2}g^{\sigma \alpha }\left( \frac{%
\partial g_{\mu \alpha }}{\partial x^{\nu }}+\frac{\partial g_{\nu \alpha }}{%
\partial x^{\mu }}-\frac{\partial g_{\mu \nu }}{\partial x^{\alpha }}\right)
.
\end{equation}

\noindent Using the requirement that only derivatives with respect to $z$
are non-zero leads to \cite{Rohrlich},

\begin{equation}
\Gamma _{03}^{0}=\Gamma _{30}^{0}=\frac{1}{V}\frac{\partial V}{\partial z}%
,\;\Gamma _{00}^{3}=\frac{V}{U^{2}}\frac{\partial U}{\partial z}%
,\;and,\;\Gamma _{33}^{3}=\frac{1}{U}\frac{\partial U}{\partial z}.
\end{equation}

\noindent All other Christoffel symbols are zero.

In order to solve for the metric components we first calculate the
relationship between the $R_{00}$ component of the Ricci tensor and the
derivatives of the metric, and substitute for the Christoffel symbols,

\begin{equation}
R_{00}=\frac{V}{U^{3}}\frac{\partial U}{\partial z}\frac{\partial V}{%
\partial z}-\frac{V}{U^{2}}\frac{\partial ^{2}V}{\partial z^{2}}.
\end{equation}

\noindent A similar calculation for the relationship between the component
of the Ricci tensor and the derivatives of the metric leads to,

\begin{equation}
R_{33}=\frac{1}{V}\frac{\partial ^{2}V}{\partial z^{2}}-\frac{1}{VU}\frac{%
\partial U}{\partial z}\frac{\partial V}{\partial z}.
\end{equation}

\noindent Again all other components are found to be identically zero.

Since the Ricci tensor is zero in source free space we obtain from the
expansion of the $R_{00}$ equation,

\begin{equation}
\frac{\partial U}{\partial z}\frac{\partial V}{\partial z}-U\frac{\partial
^{2}V}{\partial z^{2}}=0.
\end{equation}

\noindent The $R_{33}$ equation is identical to the $R_{00}$ equation. The
solution to these two equations is \cite{Rohrlich},

\begin{equation}
U=\frac{1}{\alpha }\frac{\partial V}{\partial z}
\end{equation}

\noindent where $\alpha $ is a constant. While this provides a relationship
between the components of the metric, the differential equation is
underdetermined, since both $U$ and $V$ are unknown.

The relationship between the metric components can also be established by
requiring that the metric be conformally flat as developed previously by
Tilbrook.\cite{Tilbrook} The restriction of conformal flatness on the metric
assumes the existence of a diffeomorfism or coordinate transformation,

\begin{equation}
t^{\prime }=t^{\prime }\left( t,z\right) ,\quad x^{\prime }=x,\quad
y^{\prime }=y,\quad and,\quad z^{\prime }=z^{\prime }\left( t,z\right) ,
\end{equation}

\noindent where the prime coordinates are a Minkowski space,

\begin{equation}
ds^{2}=dt^{\prime 2}+dx^{2}+dy^{2}+dz^{\prime 2},
\end{equation}

\noindent and the unprimed coordinates are the accelerated coordinate
system. The assumption of conformal flatness results in the same
relationship for the metric components found from the solution to the field
equations, which formally demonstrates the equivalence of a uniform plane
symmetric gravitational field and rectilinear acceleration.

The solution to the field equations for a plane symmetric field or equally
the requirement that the accelerated coordinate system is conformally flat
does not uniquely determine the components of the metric for an accelerated
coordinate system. To uniquely determine the metric components some
ancillary conditions to the physical restrictions on the metric are
required. These added restrictions on the metric components are provided by
requiring that the coordinate system satisfy some additional constraints.
These added constraints are the coordinate conditions for the metric. The
components of the metric of the accelerated coordinate system are restricted
by the relation between the time and space components $U=\frac{1}{\alpha }%
\frac{\partial V}{\partial z}$ and are not explicitly time dependent. Any
two accelerated coordinate systems that satisfy the relation between the
metric components will have a time independent coordinate transformation and
the space part of the metric will only be position dependent, $V\left(
z\right) =V\left( Z\right) .$ The coordinate time for any accelerated
coordinate system is then independent of the coordinate conditions. This is
shown by first writing the metric for some accelerated coordinate system, 
\begin{equation}
ds^{2}=-V\left( T\right) ^{2}dT^{2}+U\left( Z\right) ^{2}dZ^{2}.
\end{equation}

\noindent For some second accelerated coordinate system the spacial part of
the metric components can be equated $V\left( z\right) =V\left( Z\right) ,$
as well as the spacial part of the metric,

\begin{equation}
U\left( Z\right) dZ=\frac{1}{\alpha }\frac{\partial V\left( z\right) }{%
\partial z}\frac{dz}{dZ}\frac{dZ}{dz}dz,
\end{equation}

\noindent and the metric expressed in terms of the space part of this
coordinate system as

\begin{equation}
ds^{2}=-V\left( z\right) ^{2}dT^{2}+U\left( z\right) ^{2}dz^{2}.
\end{equation}

\noindent It follows from the invariance of the 4-space interval that the
coordinate time for all accelerated coordinate systems are the same, $%
dT^{2}=dt^{2},$ provided that the acceleration and the velocity of the
reference frames are the same.

\section{Constant Acceleration of Light}

The Einstein field equations do not uniquely determine the components of the
metric since both $U$ and $V$ are unknown in the differential equations. An
exact solution for the metric components can be obtained by requiring that
the metric produce the null line element of light. Assume for example the
null interval of light moving in the direction $z$,

\begin{equation}
0=-V^{2}dt^{2}+U^{2}dz^{2}.
\end{equation}

\noindent Now recalling that $U=\frac{1}{\alpha }\frac{\partial V}{\partial z%
}$ leads to an equation for the metric components,

\begin{equation}
0=-V^{2}dt^{2}+\frac{1}{\alpha ^{2}}\left( \frac{\partial V}{\partial z}%
\right) ^{2}dz^{2}.
\end{equation}

\noindent This equation can be rearranged and solved for the velocity,

\begin{equation}
\frac{dz}{dt}=\alpha \left( \frac{\partial \ln \left( V\right) }{\partial z}%
\right) ^{-1}.
\end{equation}

\noindent Here we will look for solutions in which the acceleration is a
constant. This is equivalent to looking for a coordinate system in which the
acceleration with respect to coordinate time $t$ is constant.
Differentiating the velocity with respect to time $t$ and setting $\frac{%
d^{2}z}{dt^{2}}=-a$ we obtain a differential equation, which has the
solution,

\begin{equation}
V=e^{-\nu }e^{\sqrt{\mu +2\frac{\alpha ^{2}}{a}z}}.
\end{equation}

\noindent The constants $\alpha ,$ $\mu ,$ and $\nu $ are determined by
requiring that the metric components become that of Minkowski coordinates in
the limit of zero acceleration and that the metric agree with the weak field
limit to first order and we have the solution for the metric,

\begin{equation}
ds^{2}=-e^{-2}e^{2\sqrt{1+2az}}dt^{2}+dx^{2}+dy^{2}+\frac{e^{-2}e^{2\sqrt{%
1+2az}}}{1+2az}dz^{2}.
\end{equation}

\noindent The velocity of light in the accelerated coordinate system can be
calculated by setting the 4-space interval to zero , $\frac{dz}{dt}=\sqrt{%
1+2az},$ and the coordinate velocity of light is found to be position
dependent in the accelerated coordinate system.

\section{Lass Metric}

Another unique solution can be obtained by choosing a coordinate system that
satisfies the harmonic coordinate conditions (HCC). The restrictions on the
metric components required to satisfy the HCC can be written in the form 
\cite{Weinberg},

\begin{equation}
\frac{\partial \sqrt{g}g^{\lambda \alpha }}{\partial x^{\alpha }}=0.
\end{equation}

\noindent where $g=-Det\left( g_{\mu \nu }\right) $. The metric components
are functions of $z$ only and the metric is also diagonal which insures that
the only nonzero terms are $\lambda =3$ and also noting that for a diagonal
metric $g^{33}=-g_{33}$,

\begin{equation}
\frac{\partial }{\partial \xi }\left( \frac{\sqrt{g}}{g_{33}}\right) =0.
\end{equation}

\noindent This expression can be integrated once, with $C$ a constant,
solving the differential equation,

\begin{equation}
V\left( \xi \right) =e^{C\xi }.
\end{equation}

\noindent Matching the weak field limit $C=a$ and noting that locally $%
g_{33}\left( \xi =0\right) =1\ $requiring $\alpha =a$ uniquely determines
the metric,

\begin{equation}
ds^{2}=-e^{2a\xi }dt^{2}+dx^{2}+dy^{2}+e^{2a\xi }d\xi ^{2}.
\end{equation}

\noindent Note that for light $\frac{d\xi }{dt}=1$ and $\frac{d^{2}\xi }{%
dt^{2}}=0$. This is the same property as a photon in a special relativistic
coordinate system moving with a constant velocity. The assumption of the HCC
leads to a zero coordinate acceleration for light which is consistent with a
momentarily co-moving frame (MCMF). The coordinate transformation from the
MCMF to the accelerated coordinate system is only a function of the spatial
coordinates,

\begin{equation}
\xi =\frac{-1+\sqrt{1+2az}}{a}.
\end{equation}

\section{Rindler Metric}

The final coordinate condition that will be considered is the requirement
that the coordinate 3-space interval is equal to the proper 3-space
interval. \cite{Misner et al}\cite{Rindler 66}\cite{Tilbrook} This
coordinate condition results in a unique solution to the field equation and
the Rindler metric. Recall that $U=\frac{1}{\alpha }\frac{\partial V}{%
\partial Z}$ and to insure that the coordinate condition is satisfied set $%
U=1.$ Consistent with the weak field limit the metric is

\begin{equation}
ds^{2}=-\left( 1+aZ\right) ^{2}dt^{2}+dx^{2}+dy^{2}+dZ^{2}.
\end{equation}

\noindent The coordinate transformation from the Rindler coordinates to MCMF
of the Lass coordinates is

\begin{equation}
\xi =\frac{\ln \left( 1+aZ\right) }{a}.
\end{equation}

\noindent In the Rindler metric the coordinate velocity of light is $\frac{dZ%
}{dt}=1+aZ$ and the coordinate acceleration of light is $\frac{d^{2}Z}{dt^{2}%
}=a+a^{2}Z.$ Both the coordinate velocity and coordinate acceleration are
position dependent in the Rindler coordinate system.

\section{Conclusion}

The field equations for the static plane symmetric field were solved and
provide a restriction on the components of the metric that is identical to
the restriction of a conformally flat metric. In order to uniquely determine
the components of the metric the appropriate coordinate conditions
restricting the metric to the accelerated coordinates, the Lass coordinates,
and the Rindler coordinates were applied to the field equations in order to
uniquely determine the components of the metric. While the three coordinate
conditions resulted in different 3-space intervals the coordinate time was
found to be the same for all three metrics and in general the same for all
accelerated coordinate systems with the same velocity and acceleration.

\end{document}